\documentclass[12pt,preprint]{aastex}

\providecommand{\etal}{et~al.}

\received{}
\revised{}
\accepted{}

\slugcomment{Accepted for Publication in {\sl AJ}}

\shorttitle{The Optical Afterglow of GRB~011211}
\shortauthors{S. Holland, et~al.}

\begin{document}


\title{The Optical Afterglow of the Gamma-Ray Burst
            GRB~011211\protect\footnote{Based on observations
            collected at the OGLE 1.3m telescope, the Magellan 6.5m
            Walter Baade telescope, the FLWO 1.2m telescope, and the
            Vatican Advanced Technology Telescope.}}

\author{Stephen~T.~Holland\altaffilmark{1},
        I. Soszy{\'n}ski\altaffilmark{2},
        Michael~D.~Gladders\altaffilmark{3},
        L. F. Barrientos\altaffilmark{4},
        P. Berlind\altaffilmark{5},
        David~Bersier\altaffilmark{5},
        Peter~M.~Garnavich\altaffilmark{1},
        Saurabh~Jha\altaffilmark{5},
        K.~Z.~Stanek\altaffilmark{5}}

\altaffiltext{1}{Department of Physics,
                 University of Notre Dame,
                 Notre Dame, IN 46556--5670,
                 U.S.A.
                 \email{sholland@nd.edu,
                        pgarnavi@miranda.phys.nd.edu}}

\altaffiltext{2}{Warsaw University Observatory,
                 Al.\ Ujazdowskie 4,
                 00--478 Warszawa,
                 Poland
                 \email{soszynski@astrouw.edu.pl}}

\altaffiltext{3}{Carnegie Observatories,
                 813 Santa Barbara Street,
                 Pasadena, CA 91101--1292,
                 U.S.A.
                 \email{gladders@ociw.edu}}

\altaffiltext{4}{Department of Astronomy \& Astrophysics,
                 Pontificia Universidad Cat{\'o}lica de Chile,
                 Casilla 306, Santiago, 22,
                 Chile
                 \email{barrientos@astro.puc.cl}}

\altaffiltext{5}{Harvard--Smithsonian Center for Astrophysics,
                 60 Garden Street,
                 Cambridge, MA 02138,
                 U.S.A.
                 \email{perry@flwo60.sao.arizona.edu,
                        dbersier@m31.harvard.edu,
                        saurabh@cfa.harvard.edu,
                        kstanek@cfa.harvard.edu}}


\begin{abstract}

     We present early-time optical photometry and spectroscopy of the
optical afterglow of the gamma-ray burst \objectname{GRB~011211}.  The
spectrum of the optical afterglow contains several narrow metal lines
which are consistent with the burst occurring at a redshift of $2.140
\pm 0.001$.  The optical afterglow decays as a power law with a slope
of $\alpha = 0.83 \pm 0.04$ for the first $\approx 2$ days after the
burst at which time there is evidence for a break.  The slope after
the break is $\ge 1.4$.  There is evidence for rapid variations in the
$R$-band light approximately 0.5 days after the burst.  These
variations suggest that there are density fluctuations near the
gamma-ray burst on spatial scales of approximately 40--125 AU\@. The
magnitude of the break in the light curve, the spectral slope, and the
rate of decay in the optical, suggest that the burst expanded into an
ambient medium that is homogeneous on large scales.  We estimate that
the local particle density is between approximately 0.1 and 10
cm$^{-3}$ and that the total gamma-ray energy in the burst was
$1.2$--$1.9 \times 10^{50}$ erg.  This energy is smaller than, but
consistent with, the ``standard'' value of $(5 \pm 2) \times 10^{50}$
erg.  Comparing the observed color of the optical afterglow with
predictions of the standard beaming model suggests that the rest-frame
$V$-band extinction in the host galaxy is $\lesssim 0.03$ mag.

\end{abstract}




\section{Introduction\label{SECTION:intro}}

     The gamma-ray burst (GRB) \objectname{GRB~011211} was detected in
the constellation Crater by the {\sl BeppoSAX\/} satellite at 19:09:21
UT on 2001 Dec.\ 11.  The burst was a shallow, long event with two
peaks and a total duration of approximately 270 s making it the
longest event that has been localized by {\sl BeppoSAX}.  The temporal
profiles of the event were similar in gamma- and $X$-rays.  {\sl
BeppoSAX\/} measured a gamma-ray fluence of $5 \times 10^{-6}$ erg
cm$^{-2}$ between 40 keV and 700 keV \citep{FAG2002}.  Approximately
ten hours after the burst occurred \citet{GHP2001} identified an
optical source within the {\sl BeppoSAX\/} error circle that was not
present in the Digital Sky Survey~2.  \citet{BB2001} and
\citet{JPH2001} reported that this source was fading and
\citet{SHG2001} estimated that it had a power law decay with a slope
of $\alpha = 0.93 \pm 0.06$.  \citet{FVR2001} found, and
\citet{GHG2001} confirmed, a redshift of $z = 2.14$ based on several
absorption lines in the spectrum of the optical afterglow (OA).
\citet{BRF2001} identified a host galaxy with $R_\mathrm{host} = 25.0
\pm 0.3$ and found that the OA is offset $0\farcs5$ southeast from the
center of this object.

     In this paper we present photometry and spectroscopy of the OA of
\objectname{GRB~011211} taken between $\approx 0.5$ and 2.7 days after
the burst occurred.  Our data suggests that there were rapid
fluctuations in the optical flux approximately 12 hours after the
burst occurred.  We adopt a cosmology with a Hubble parameter of $H_0 =
65$ km s$^{-1}$ Mpc$^{-1}$, a matter density of $\Omega_m = 0.3$, and
a cosmological constant of $\Omega_\Lambda = 0.7$.  For this cosmology
a redshift of $z = 2.140$ corresponds to a luminosity distance of
18.18 Gpc and a distance modulus of 46.30.  One arcsecond corresponds
to 28.06 comoving kpc, or 8.94 proper kpc.  The lookback time is 11.28
Gyr.


\section{The Photometric Data\label{SECTION:phot}}

     The OA for \objectname{GRB~011211} is located at R.A.\ $=$
11:15:17.98, Dec.\ $=$ $-$21:56:56.2 (J2000)~\citep{GHP2001}, which
corresponds to Galactic coordinates of
$(b^\mathrm{II},\ell^\mathrm{II}) = (+35\fdg695,333\fdg491)$.  The
reddening maps of \citet{SFD1998} give a Galactic reddening of
$E_{B\!-\!V} = 0.043 \pm 0.020$ mag in this direction.  The
corresponding Galactic extinctions are $A_V = 0.142$, $A_R = 0.114$,
and $I = 0.083$.  We estimate the extinction in the host galaxy in
\S~\ref{SECTION:extinct}.

\begin{figure}
\plotone{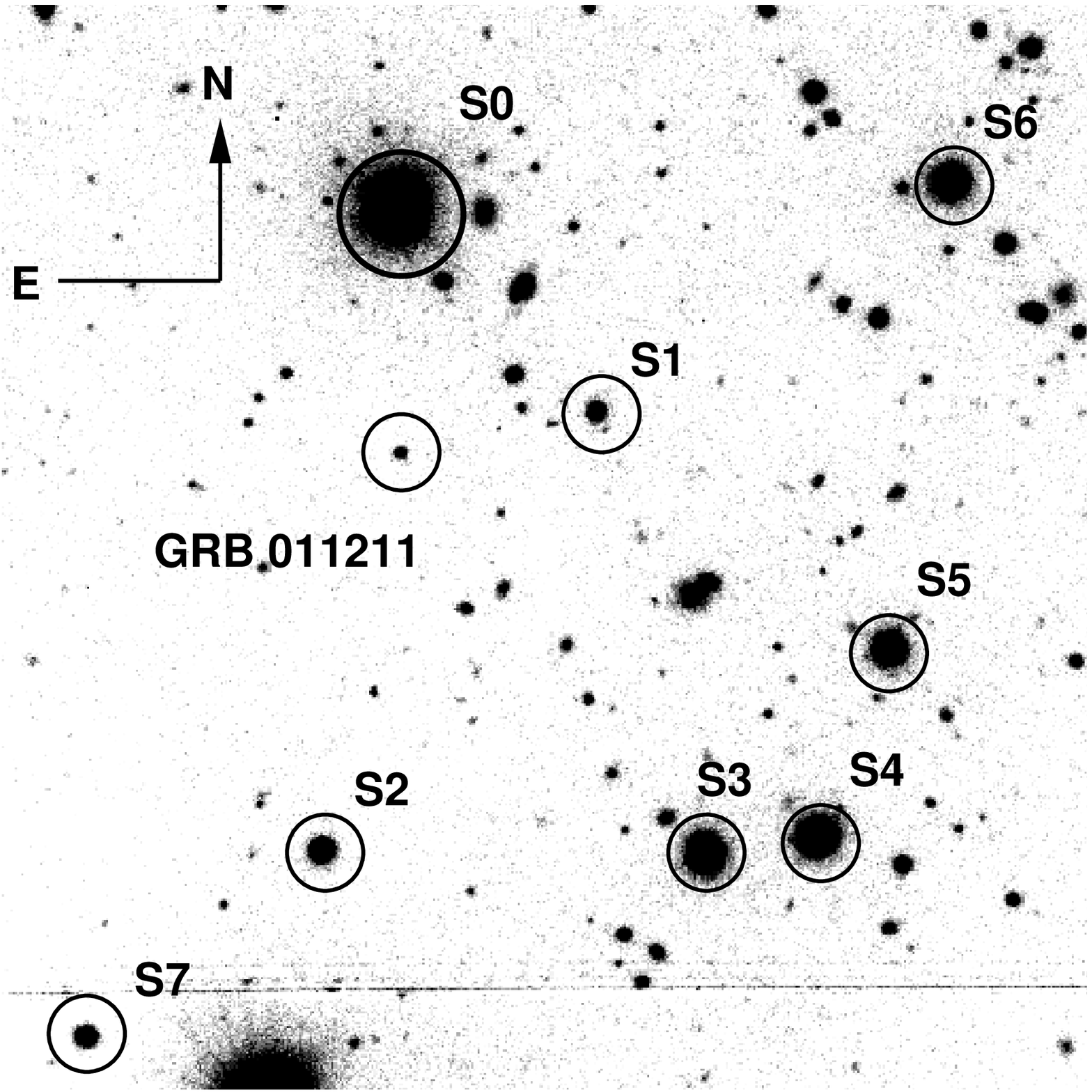}
\caption{This Figure is a combined $R$-band image of the field of
\protect\objectname{GRB~011211}.  The OA, and the stars used for
calibration (see Table~\ref{TABLE:standards}), are circled.  The
USNO--A2.0 star U0675\_11427359 \citep{H2001,H2002} is our star S2.
Each line in the compass is $30\arcsec$ long.  The horizontal streaks
near the bottom of the image are due to a bad column on the OGLE-III
CCD.}
\label{FIGURE:finder}
\end{figure}

     Our photometry data was collected at three telescopes:\ the
Optical Gravitational Lensing Experiment (OGLE) 1.3m telescope at the
Las Campanas Observatory, the Magellan 6.5m Walter Baade telescope at
Las Campanas, and the Fred Lawrence Whipple Observatory (FLWO) 1.2m
telescope.  The OGLE 1.3m telescope was equipped with the 8k $\times$
8k OGLE-III CCD, which a had a scale of $0\farcs26$ per pixel, a gain
setting of 1.0 e$^{-}$/ADU, and a read-out noise of 5 e$^-$.  The
Magellan images were taken with the LDSS-2 imaging spectrograph in its
imaging mode.  It had a scale of $0\farcs378$ per pixel, a gain of 1.0
e$^-$/ADU, and a read-out noise of 7 e$^-$.  The FLWO images were
taken with the ``4Shooter'' CCD mosaic \citep{S2002}.  The pixel scale
was $0\farcs67$ (binned $\times 2$), the gain was 4.0 e$^-$/ADU, and
the read-out noise was 10 e$^-$.  A log of the observations is given
in Table~\ref{TABLE:phot}.  Fig.~\ref{FIGURE:finder} shows the field
containing the OA\@.

\begin{deluxetable}{cccccccc}
\tabletypesize{\footnotesize}
\tablewidth{0pt}
\tablecaption{The photometry.\label{TABLE:phot}}
\tablehead{%
	\colhead{UT Date} &
	\colhead{JD$-$2450000} &
	\colhead{$t$} & 
	\colhead{Telescope} &
	\colhead{Filter} &
	\colhead{Magnitude} &
	\colhead{Seeing ($\arcsec$)} &
	\colhead{Exposure (s)}
}
\startdata
Dec.\ 12.3028 & 2255.8063 & 0.5081 & OGLE  & $V$ & $20.69 \pm 0.05$ & $1\farcs43$ & 600 \\
Dec.\ 13.2668 & 2256.7670 & 1.4688 & OGLE  & $V$ & $21.59 \pm 0.08$ & $1\farcs32$ & 600 \\
Dec.\ 14.4844 & 2257.9844 & 2.6862 & FLWO  & $V$ & $>22.3 \pm 0.2$  & $1\farcs72$ & 2 $\times$ 600 \\

Dec.\ 12.2482 & 2255.7482 & 0.4500 & OGLE  & $R$ & $20.09 \pm 0.07$ & $1\farcs67$ & 180 \\
Dec.\ 12.2500 & 2255.7500 & 0.4518 & Baade & $R$ & $20.28 \pm 0.02$ & $1\farcs23$ &  60 \\
Dec.\ 12.2567 & 2255.7567 & 0.4585 & OGLE  & $R$ & $20.40 \pm 0.04$ & $1\farcs70$ & 600 \\
Dec.\ 12.2771 & 2255.7771 & 0.4789 & OGLE  & $R$ & $20.37 \pm 0.04$ & $1\farcs58$ & 600 \\
Dec.\ 12.2945 & 2255.7945 & 0.4963 & OGLE  & $R$ & $20.25 \pm 0.03$ & $1\farcs43$ & 600 \\
Dec.\ 12.3226 & 2255.8226 & 0.5244 & OGLE  & $R$ & $20.33 \pm 0.04$ & $1\farcs48$ & 600 \\
Dec.\ 12.3400 & 2255.8400 & 0.5418 & Baade & $R$ & $20.51 \pm 0.04$ & $0\farcs83$ &  60 \\
Dec.\ 13.2418 & 2256.7418 & 1.4436 & OGLE  & $R$ & $21.48 \pm 0.11$ & $1\farcs40$ & 600 \\
Dec.\ 13.3432 & 2256.8432 & 1.5450 & OGLE  & $R$ & $21.41 \pm 0.07$ & $1\farcs16$ & 600 \\
Dec.\ 13.3147 & 2256.8147 & 1.5165 & Baade & $R$ & $21.35 \pm 0.09$ & $0\farcs74$ &  60 \\
Dec.\ 14.5132 & 2258.0132 & 2.7150 & FLWO  & $R$ & $>22.4 \pm 0.2$  & $1\farcs68$ & 2 $\times$ 900 \\

Dec.\ 12.2662 & 2255.7662 & 0.4680 & OGLE  & $I$ & $19.88 \pm 0.07$ & $1\farcs57$ & 600 \\
Dec.\ 12.2863 & 2255.7863 & 0.4881 & OGLE  & $I$ & $19.90 \pm 0.06$ & $1\farcs38$ & 600 \\
Dec.\ 12.3145 & 2255.8145 & 0.5163 & OGLE  & $I$ & $19.91 \pm 0.05$ & $1\farcs40$ & 600 \\
Dec.\ 12.3307 & 2255.8307 & 0.5325 & OGLE  & $I$ & $20.06 \pm 0.05$ & $1\farcs31$ & 600 \\
Dec.\ 13.2559 & 2256.7559 & 1.4577 & OGLE  & $I$ & $20.78 \pm 0.11$ & $1\farcs24$ & 600 \\
Dec.\ 13.3533 & 2256.8533 & 1.5551 & OGLE  & $I$ & $20.86 \pm 0.11$ & $1\farcs17$ & 600 \\
Dec.\ 14.5153 & 2258.0153 & 2.7171 & FLWO  & $I$ & $>21.0 \pm 0.2$  & $1\farcs98$ & 2 $\times$ 600 \\
\enddata
\tablecomments{No corrections for extinction have been applied to the
photometry in this Table.}
\end{deluxetable}

     The field stars around \objectname{GRB~011211} were calibrated on
2002 Jan.\ 12 (UT) using images taken with the Vatican Advanced
Technology Telescope.  Landolt standard fields \citep{L1992} were
observed in $U\!BV\!RI$ filters throughout the night under photometric
conditions.  We derived airmass and color corrections and computed
magnitudes for eight stars near \objectname{GRB~011211}.  These
magnitudes and colors are given in Table~\ref{TABLE:standards}.  The
magnitudes for the USNO-A2 star \objectname{U0675\_11427359} (our star
S2) are consistent with those of \citet{H2002}.

\begin{deluxetable}{ccclcccc}
\tabletypesize{\scriptsize}
\tablewidth{0pt}
\tablecaption{Field Standards.\label{TABLE:standards}}
\tablehead{%
	\colhead{Star} &
	\colhead{R.A.(J2000)} &
	\colhead{Dec.(J2000)} &
	\colhead{$V$} &
	\colhead{$U\!-\!B$} & 
	\colhead{$B\!-\!V$} & 
	\colhead{$V\!-\!R$} & 
	\colhead{$V\!-\!I$}
}
\startdata
S0 & 11:15:18.02 & $-$21:56:13.2 & $14.42 \pm 0.02$ &  $0.057 \pm 0.036$ & $0.604 \pm 0.028$ & $0.353 \pm 0.028$ & $0.684 \pm 0.028$ \\
S1 & 11:15:15.52 & $-$21:56:49.4 & $19.75 \pm 0.05$ &  \nodata           & $1.181 \pm 0.071$ & $0.779 \pm 0.071$ & $1.484 \pm 0.071$ \\
S2 & 11:15:19.02 & $-$21:58:05.1 & $18.01 \pm 0.03$ & $-0.204 \pm 0.050$ & $0.481 \pm 0.042$ & $0.326 \pm 0.042$ & $0.650 \pm 0.042$  \\
S3 & 11:15:14.21 & $-$21:58:06.4 & $16.42 \pm 0.03$ &  $0.580 \pm 0.050$ & $0.904 \pm 0.042$ & $0.556 \pm 0.042$ & $1.100 \pm 0.042$ \\
S4 & 11:15:12.82 & $-$21:58:03.6 & $16.51 \pm 0.03$ &  $0.966 \pm 0.050$ & $1.013 \pm 0.042$ & $0.653 \pm 0.042$ & $1.165 \pm 0.042$ \\
S5 & 11:15:11.88 & $-$21:57:31.2 & $17.52 \pm 0.04$ &  $1.181 \pm 0.064$ & $1.517 \pm 0.057$ & $1.053 \pm 0.057$ & $2.484 \pm 0.057$ \\
S6 & 11:15:11.05 & $-$21:56:10.0 & $16.49 \pm 0.03$ &  $0.078 \pm 0.050$ & $0.654 \pm 0.042$ & $0.387 \pm 0.042$ & $0.745 \pm 0.042$ \\
S7 & 11:15:21.96 & $-$21:58:36.6 & $19.45 \pm 0.05$ &  \nodata           & $1.476 \pm 0.071$ & $1.012 \pm 0.071$ & $2.314 \pm 0.071$ \\
\enddata
\tablecomments{These stars are identified in Fig.~\ref{FIGURE:finder}.}
\end{deluxetable}

     We used {\sc DAOPhot II}~\citep{S1987} and {\sc AllStar}
\citep{SH1988} to perform point-spread function fitting photometry on
the calibrated field stars, and on the OA, on each image.  A
zero-point offset was calculated for each filter on each night and
applied to the observed magnitudes.  Our calibrated photometry for the
OA is presented in Table~\ref{TABLE:phot}.  The two FLWO images in
each band were averaged to obtain the final images.  The OA was not
detected in any of our 2001 Dec.\ 14 FLWO images.  Therefore, we
estimated upper limits to the brightness of the OA on 2001 Dec.\ 14.5
from the limiting magnitudes of these images, which are listed in
Table~\ref{TABLE:phot}.

     The OA is consistent with a point source in all of our images.
We aligned and co-added all of our $R$-band images to obtain a deep
image (shown in Fig.~\ref{FIGURE:finder}) with a limiting magnitude of
$R_{\lim} \approx 23.5 \pm 0.2$.  There is no evidence for a host
galaxy in this image.


\section{The Spectral Data\label{SECTION:spectra}}

     A spectrum of the OA was obtained with the Magellan 6.5m Walter
Baade telescope using the LDSS-2 imaging spectrograph on 2001 Dec.\
13.3 UT, approximately 1.5 days after the burst.  The slit width was
$1\arcsec$ and the total exposure time was $4 \times 620$ s.  The
resolution is $12$ {\AA}.  The spectra were reduced in the standard
manner and the wavelength calibration was done using a He--Ne arc.
Several weak metal lines were identified to determine the redshift of
the system.  These are listed in Table~\ref{TABLE:lines}.  The mean
redshift is $z = 2.140 \pm 0.001$ (standard error [se]).
Table~\ref{TABLE:lines} also lists the observed and rest-frame
equivalent widths for each line.

\begin{figure}
\plotone{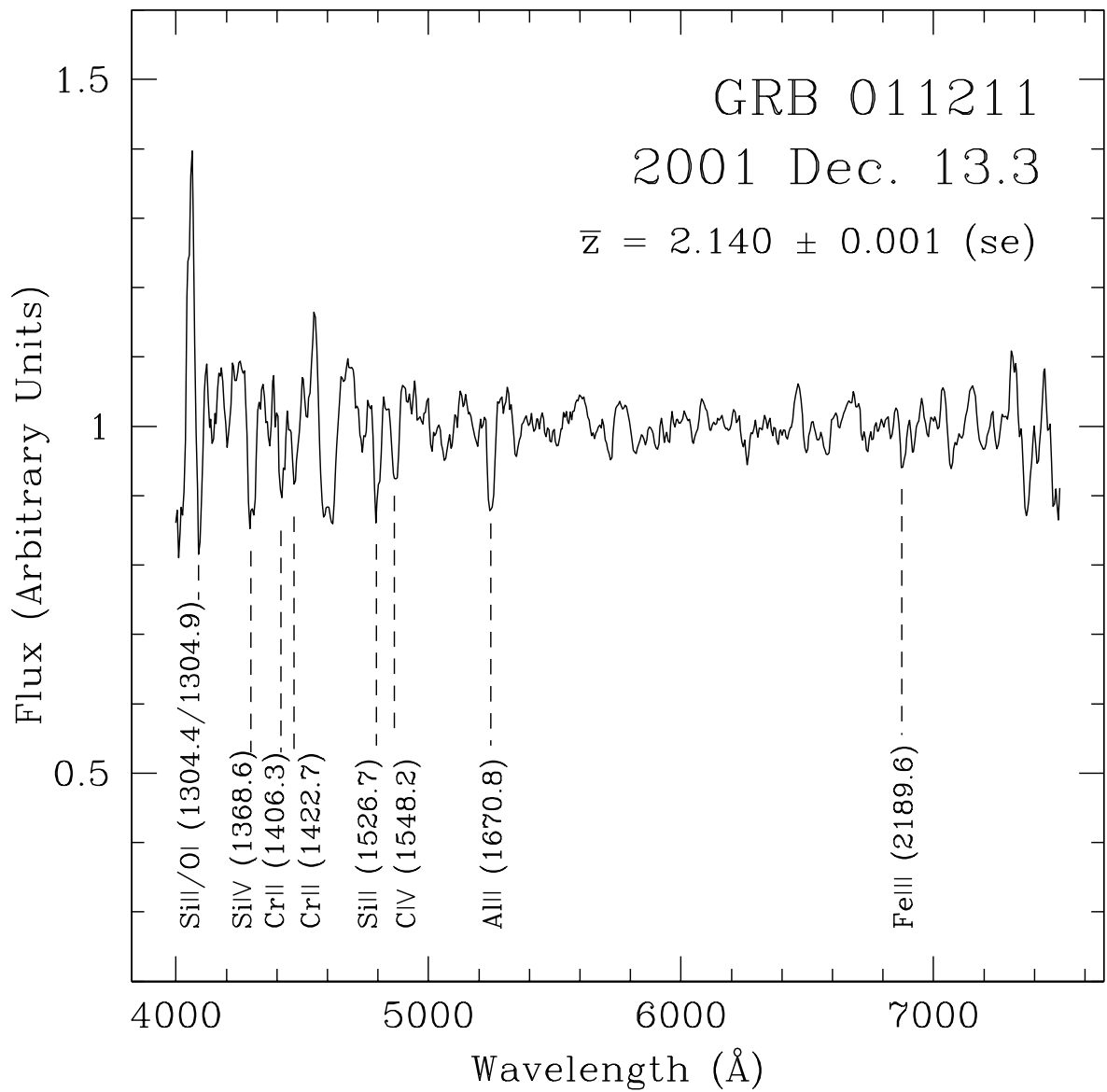}
\caption{This Figure shows the Magellan/LDSS-2 spectrum of
\protect\objectname{GRB~011211} taken on 2001 Dec.\ 13.3.  The
spectrum was smoothed with a 30 {\AA} boxcar to make the metal lines
more clear and normalized to the continuum.  The absorbtion lines
listed in Table~\ref{TABLE:lines} are marked and their rest-frame
wavelengths (in {\AA}) are given in brackets.  There is an
unidentified broad feature at $\approx 4600$ {\AA}.  The mean redshift
from the identified lines is $z = 2.140 \pm 0.001$ (se).}
\label{FIGURE:spectrum}
\end{figure}

     The measured absorption line widths for \objectname{GRB~011121}
are quite similar to those observed in \objectname{GRB~000301C}
\citep{JFG2001} but weaker than the exceptional case of
\objectname{GRB~000926} \citep{CGH2001,FGM2002}.  Both of these GRBs
have redshifts similar to that of \objectname{GRB~011211}.  We also
observe an unidentified, broad feature at approximately 4600 \AA\@.
This may be an additional absorption system along the line of sight,
but the signal-to-noise ratio of our spectrum is not high enough to
confirm this hypothesis.  A plot of the spectrum, with our line
identifications, is given in Fig.~\ref{FIGURE:spectrum}.

\begin{deluxetable}{lccccc}
\tabletypesize{\footnotesize}
\tablewidth{0pt}
\tablecaption{Spectral Lines.\label{TABLE:lines}}
\tablehead{%
	\colhead{Line} &
	\colhead{$\lambda$ (\AA)} &
	\colhead{$\lambda_0$ (\AA)} & 
	\colhead{$z$} & 
	\colhead{EW (\AA)} & 
	\colhead{EW$_0$ (\AA)}
}
\startdata
\ion{Si}{2}/ & 4090.2 & 1304.4 & 2.1358 & 11.28 & 3.59 \\
\ion{O}{1}   &        & 1304.9 & 2.1346 &       &      \\
\ion{Si}{4}  & 4296.1 & 1368.1 & 2.1390 &  5.41 & 1.72 \\
\ion{Cr}{2}  & 4415.6 & 1406.3 & 2.1399 &  2.17 & 0.69 \\
\ion{Cr}{2}  & 4467.0 & 1422.7 & 2.1398 &  2.84 & 0.90 \\
\ion{Si}{2}  & 4794.0 & 1526.7 & 2.1401 &  4.17 & 1.33 \\
\ion{C}{4}   & 4865.3 & 1548.2 & 2.1426 &  4.08 & 1.30 \\
\ion{Al}{2}  & 5247.8 & 1670.8 & 2.1409 &  5.11 & 1.63 \\
\ion{Fe}{3}  & 6875.2 & 2189.6 & 2.1399 &  3.42 & 1.09 \\
\enddata
\end{deluxetable}

\section{The Optical Light Curve\label{SECTION:light_curve}}

\begin{figure}
\plotone{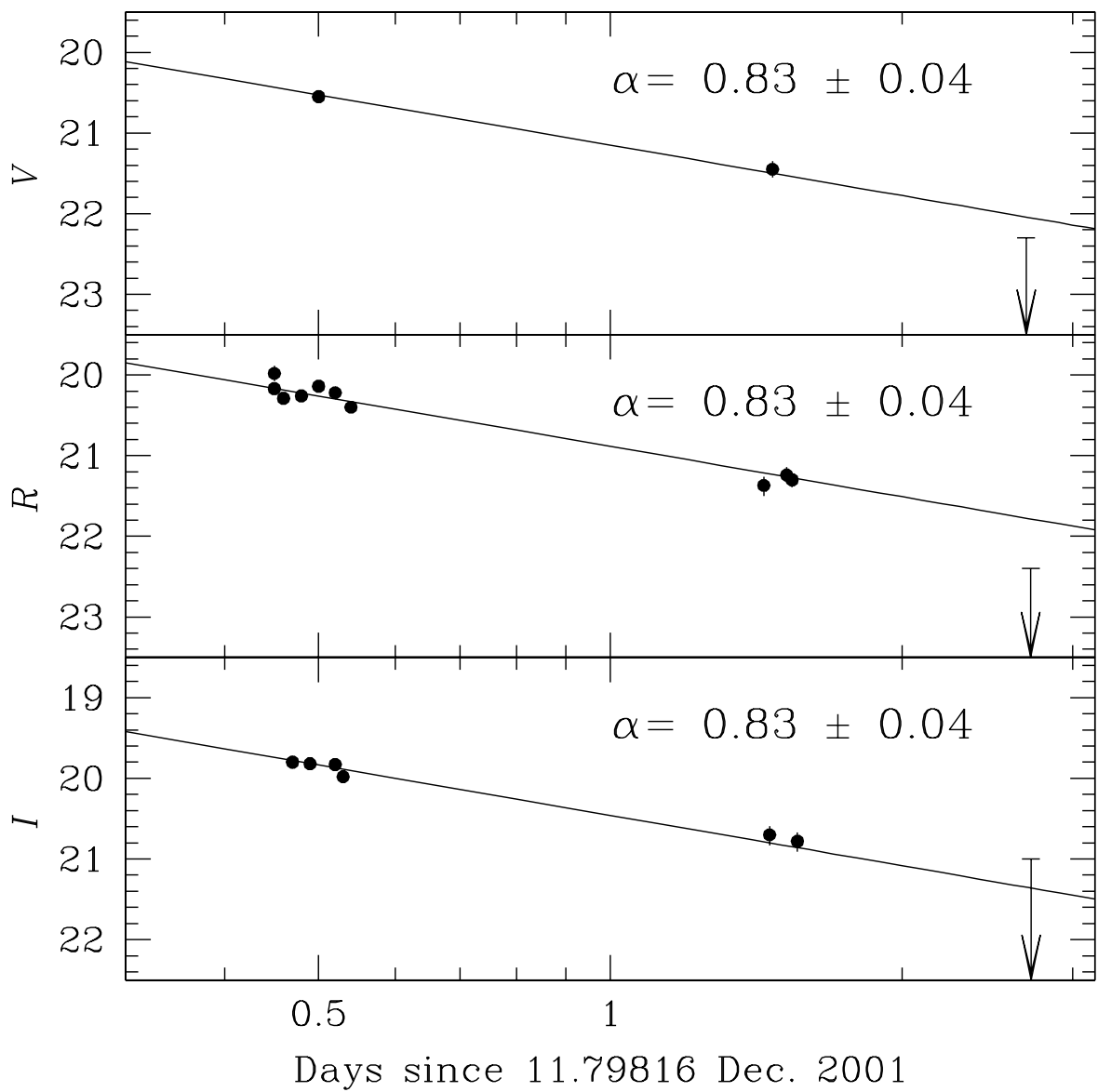}
\caption{This Figure shows the best-fitting power law to our $V\!RI$
observations of the OA of \protect\objectname{GRB~011211}.  The slope
is consistent with the early-time slopes seen for other GRB OAs.  The
non-detection of the OA on 2001 Dec.\ 14.5 (represented by the arrows)
suggests that the break occurred between 1.6 and 2.7 days after the
burst.  The photometry has been corrected for Galactic extinction but
not for extinction in the host.}
\label{FIGURE:fit}
\end{figure}

     The $V\!RI$ light curves of the OA of \objectname{GRB~011211} are
shown in Fig.~\ref{FIGURE:fit}.  The flux from the host galaxy
\citep{BRF2001} has not been subtracted.  The pixel sizes in our
images are comparable to the reported offset between the OA and the
host ($0\farcs5$), so our quoted magnitudes include the flux from both
the OA and the host galaxy.  However, the host is faint enough that it
will contribute only $\approx 1$\% of the flux when the OA has $R
\approx 20$ and $\approx 10$\% of the flux when the OA has $R \approx
22$.  This corresponds to systematic errors of only 0.01--0.10 mag.
Therefore, we believe that the flux from the host galaxy does not
significantly affect our results.

     We fit all of the data in Table~\ref{TABLE:phot} up to 1.6 days
after the burst with a power law of the form

\begin{equation}
M_j = K_j - 2.5\log_{10}(C_j t^{-\alpha})
\label{EQUATION:powerlaw}
\end{equation}

\noindent
where $M_j$ is the magnitude for filter $j$, $K_j$ is the photometric
zero point for filter $j$ taken from \citet{FSI1995}, $t$ is the time
in days since the burst occurred, $\alpha$ is the slope, and $C_j$ is
a normalization constant representing the flux from the OA in filter
$j$ one day after the burst.  The best-fitting power law to our data
has $\alpha = 0.83 \pm 0.04$ with $\chi^2/\mathrm{DOF} = 19.222/14$
and a root-mean-square (RMS) residual of 0.09 mag.

     Our power-law fit suggests that the OA should have $R = 21.9 \pm
0.1$ at $t = 2.72$ days.  However, we find that the OA has $R > 22.4$
at this time, $\approx 5 \sigma$ fainter than the predicted magnitude.
This non-detection suggests that a break occurred in the light curve
between 1.52 and 2.72 days after the burst.  If the break occurred at
more than 1.52 days then the late-time decay has a slope of $\alpha_2
\ge 1.4$, which corresponds to the decay becomeing steeper by $\Delta
\alpha = \alpha_2 - \alpha \ge 0.6$.  The FLWO $V$-band data are
consistent with this, but the FLWO $I$-band data are consistent with
there being no break in the light curve.  \citet{BRF2001} found $R =
24.8 \pm 0.3$ for the OA ten days after the burst.  If we fit a broken
power law to our $R$-band data and their point we find a late-time
slope of $\alpha_2 = 1.70 \pm 0.26$ and a break time of $t_b = 1.80
\pm 0.62$ days, which are consistent with the estimates of the
late-time slope and break time made using only our data.

     Fig.~\ref{FIGURE:fit} shows that a single power law with a slope
of $0.83 \pm 0.04$ is a good fit to the $V\!RI$ data up to 1.52 days
after the burst.  Using Eq.~\ref{EQUATION:powerlaw} we find colors of
$V\!-\!R = 0.27 \pm 0.13$, $V\!-\!I = 0.70 \pm 0.13$, and $R\!-\!I =
0.43 \pm 0.13$ before the break.  We stress that these colors are
corrected for Galactic extinction, but {\sl not\/} for extinction in
the host galaxy of \objectname{GRB~011211}.  In
\S~ref{SECTION:extinct} we show that the extinction in the host along
the line of sight to the burst is probably small compared to the
Galactic extinction in that direction.  Assuming a spectrum of the
form $f_\nu \propto \nu^{-\beta}$ we find a weighted mean dereddened
spectral slope from these colors of $\beta = 0.61 \pm 0.15$ (se).
This is similar to the dereddened spectral slope found for many other
GRBs.


\section{The Energy in the Burst\label{SECTION:energy}}

     The {\sl BeppoSAX\/} fluence is $5 \times 10^{-6}$ erg cm$^{-2}$
in the 40--700 keV band \citep{FAG2002}.  Applying a cosmological $k$
correction \citep{BFS2001} this fluence corresponds to an isotropic
equivalent energy of $E_\mathrm{iso} = 7 \times 10^{52}$ erg between
20 keV and 2000 keV.  Our $k$ correction assumes that the gamma-ray
spectrum has the same form as the \citet{BMF1993} spectrum with
$\alpha = -1$, $\beta = -2$, and $E_0 = 150$ keV.  These are the
cannonical values for a GRB's gamma-ray spectrum but \citet{BMF1993}
found that they are not universal values and that individual bursts
can have very different spectral shapes.  We estimated the uncertainty
in our $k$ correction to be $\approx 20$\% based on changing the
spectral shape parameters by factors of two.

     The opening angle, $\theta_0$, of a GRB jet is related to the
time of the break in the light curve \citep{R1999,SPH1999}.
\citet{FKS2001} cast this relation as

\begin{equation}
\theta_0 = 0.057
           {\left( t_b \over 1~\mathrm{day} \right)}^{3/8}
           {\left( 1+z \over 2 \right)}^{-3/8}
           {\left( E_\mathrm{iso} \over 10^{53}~\mathrm{erg} \right)}^{-1/8}
           {\left( \eta_\gamma \over 0.2 \right)}^{1/8}
           {\left( n \over 0.1~\mathrm{cm}^{-3} \right)}^{1/8} \mathrm{rad}
\label{EQUATION:opening_angle}
\end{equation}
where $\eta_\gamma$ is the efficiency of converting energy in the
ejecta into gamma rays and $n$ is the circumburst particle density.
\citet{FWK2000} and \citet{PK2001} find $0.001 \lesssim n \lesssim 3$
cm$^{-3}$ in the vicinity of five GRBs with a median number density of
$n = 0.1$ cm$^{-3}$.  Therefore, we adopt this as the circumburst
number density.  We will assume, as did \citet{FKS2001}, that
$\eta_\gamma = 0.2$.  Eq.~\ref{EQUATION:opening_angle} gives $\theta_0
= 3\fdg4$--$4\fdg2$ for break times of $t_b = 1.5$--$2.7$ days.  From
this we estimate that \objectname{GRB~011211}'s total beamed energy in
gamma rays, after correcting for the beam geometry, was $E_\gamma
\approx 1.2$--$1.9 \times 10^{50}$ erg.  This is only $\approx 2
\sigma$ smaller than the ``standard'' total beamed energy in gamma
rays of $5 \times 10^{50}$ erg \citep{FKS2001,PKP2001,PK2002}.  This
agreement suggests that our assumptions of $\eta_\gamma \approx 0.2$
and $n \approx 0.1$ cm$^{-3}$ are reasonable.

     In order for $E_\gamma$ to equal the ``standard'' value the
opening angle of the beam needs to be $\approx 7\degr$.  This
corresponds to a break time of $\approx 10$ days, which is
inconsistent with the observed brightness of the OA on 14 Dec. 2001.
An opening angle of $7\degr$ can be made consistent with our estimate
of the break time by increasing the circumburst particle density to $n
\approx 5$--30 cm$^{-3}$.  This is at the high end of the range of
particle densities found by \citet{FWK2000} and \citet{PK2001} for
several GRBs, but not inconsistent with there results.  Therefore, we
believe that the environment of \objectname{GRB~011211} was similar to
the environments of other GRBs.


\section{The Ambient Medium Near the Burst\label{SECTION:medium}}

     For a collimated outflow into an ambient medium with a number
density distribution of the form $n(r) \propto r^{-\delta}$
\citep{PMR1998,MRW1998} the power-law index, $\delta$, is related to
the observed magnitude of the break in the light curve by $\delta =
(4\Delta\alpha - 3) / (\Delta\alpha - 1)$.  We find $\Delta\alpha \ge
0.6$, which corresponds to $\delta \le 1.5$.  This is not consistent
with expansion into an ambient medium that is dominated by a
pre-existing stellar wind ($\delta = 2$), but might be consistent with
expansion into a homogeneous medium ($\delta = 0$).  If the ambient
medium is, on average, uniform then the magnitude of the break in the
light curve should be $\Delta\alpha = 3/4$, which implies a late-time
slope of $\alpha_2 = 1.6$.  This is consistent with the late-time
slope that we find ($\ge 1.4$) from our non-detection of the OA on
2001 Dec.\ 14.  We note that it is also in good agreement with the
late-time slope that we find ($1.70 \pm 0.26$) using the data of
\citet{BRF2001}.  Further evidence that the ambient medium is not
dominated by a pre-existing stellar wind comes from the relationships
between the time and spectral evolutions of the flux in a homogeneous
medium \citep{SPH1999} and in a wind \citep{CL1999}.  Using the
observed pre-break slope of the light curve a homogeneous medium
predicts $\beta = 0.55 \pm 0.03$ for $\delta = 0$ in the fast cooling
regime, which is within $0.5 \sigma$ of the observed value.  In the
slow cooling regime the predicted slope is $\beta = 0.22 \pm 0.03$,
which is a worse agreement.  A pre-existing wind predicts $\beta =
0.89 \pm 0.03$ for the fast cooling case and $\beta = 0.22 \pm 0.03$,
for the slow cooling case.  Both are worse fits than the fast-cooling
homogeneous medium model (although all four cases are within $2.5
\sigma$ of the observed value).  Therefore, we believe that a
homogeneous medium with fast-cooling electrons provides better
agreement to the observations than the wind model does.  The electron
power law distribution index for this case is $p = 2.1 \pm 0.1$, which
is consistent with what is seen in other bursts.

\begin{figure}
\plotone{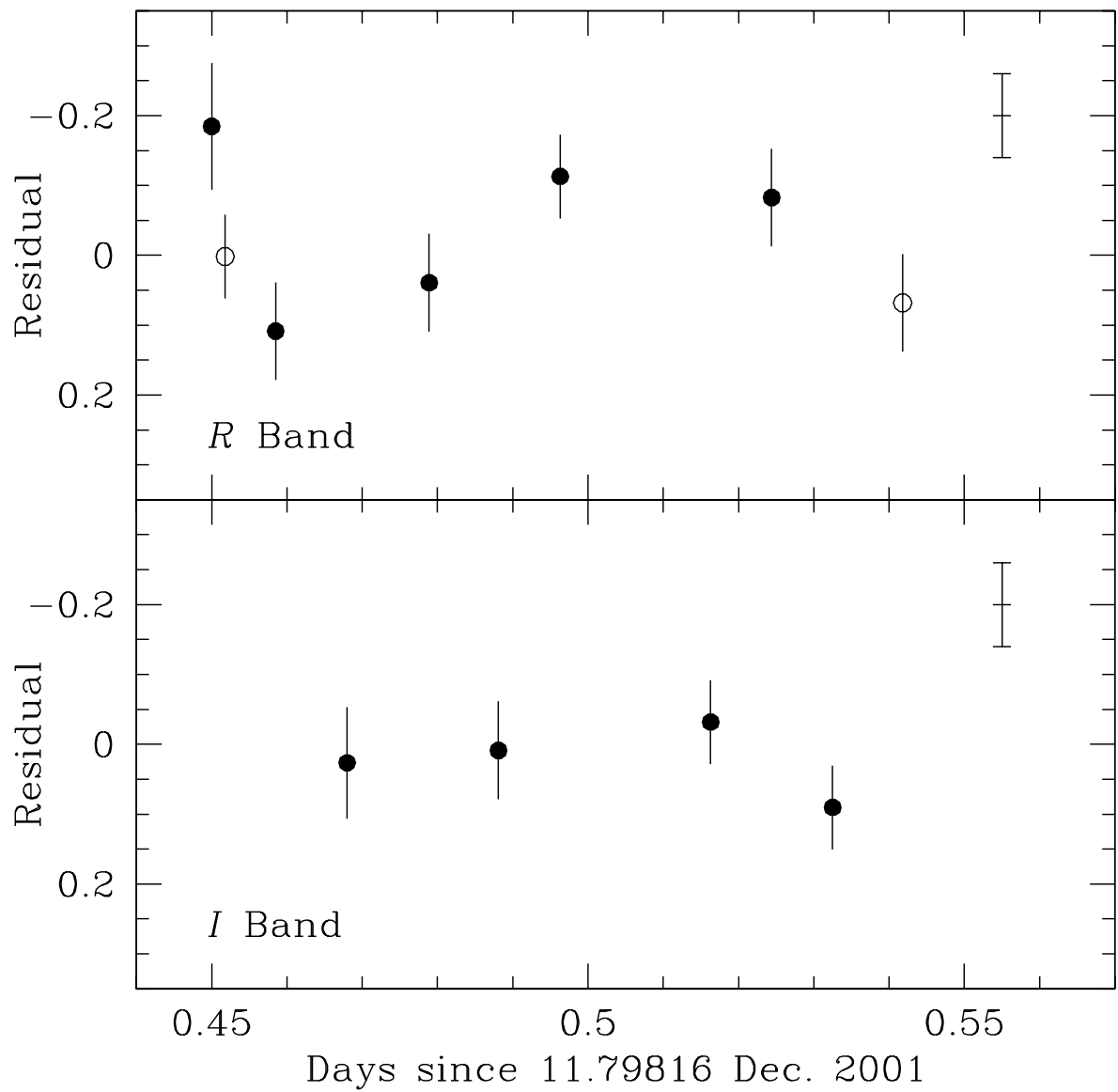}
\caption{This Figure shows our $R$- and $I$-band photometry on 2001
Dec.\ 12.  The best-fitting power law has been subtracted from the
data so that the residuals are observed magnitudes minus the power-law
fit.  Filled circles represent data taken with the OGLE 1.3m telescope
while open circles are data taken with the Magellan 6.5m Walter Baade
telescope.  The error bar in the upper right of each panel represents
the mean RMS scatter in the observed magnitudes of several field stars
within 0.2 mag of the OA on the six ($R$ band) and four ($I$) band
frames taken on this night.  The mean RMS scatter in the $R$ band is
0.06 mag for 8 stars.  In the $I$ band it is 0.06 mag for 14 stars.}
\label{FIGURE:vary}
\end{figure}

     Fig.~\ref{FIGURE:vary} suggests that there may be rapid
variations in the $R$-band light $\approx 0.5$ days after the burst.
To test this we subtracted the best-fitting power-law from the
$R$-band magnitudes and computed the residuals.  The $R$-band
residuals have a weighted mean of $-0.02 \pm 0.03$ with an RMS of 0.10
mag. This says that we can reject the hypothesis that the residuals
are consistent with random scatter about a power-law decay at the 95\%
confidence level.  Therefore, we believe that the small-scale
variations in the $R$-band light $\approx 0.5$ days after the burst
may be real.  We find no evidence for variability in the $I$-band
data.

     \citet{GLS2000} have postulated that the rapid variability seen
in \objectname{GRB~000301C} was due to gravitational microlensing.
However the rapid $R$-band variations seen in \objectname{GRB~011211}
are much smaller than those seen in the OA of
\objectname{GRB~000301C}.  An alternate explanation is that there is
small-scale structure in the local medium around the burst.
\citet{WL2000} find that linear density variations on spatial scales
of $\approx 1$--$10^3$ AU in the ambient medium that a GRB is
expanding into can cause rapid fluctuations in the optical flux
similar to those seen in Fig.~\ref{FIGURE:vary}.  Their methodology
predicts fluctuations of approximately 5\%--10\% over time periods of
$\approx 0.25$--1.50 hours for \objectname{GRB~011211}.  The observed
scatter in the $R$-band magnitudes of several field stars is 0.06 mag
and the RMS residual in the OA magnitudes is 0.10 mag.  Therefore the
RMS fractional variability in the $R$ band at about 0.5 days after the
burst is $\approx 8$\% on time-scales of $\approx 1$--2 hours.  This
is consistent with the predicted variability.

     Using Table~3 of \citet{WL2000}, and the values for the isotropic
equivalent energy of the burst and the density of the circumburst
medium that we find in Sect.~\ref{SECTION:energy}, we can estimate the
size of the density fluctuations near the burst and the radius of the
shock front.  For $E_\mathrm{iso} = 7 \times 10^{52}$ erg and $n =
0.1$--10 cm$^{-3}$ the observed $R$-band fluctuations correspond to
typical density variations on spatial scales of $\approx 40$--125
AU\@.  At 0.5 days after the burst the radius of the shock front is
$\approx 12\,000$--40\,000 AU\@1.  (The apparent faster-than-light
motion of the shock front is an illusion caused by the highly
relativistic velocity of the shock front.)  Therefore, the density
fluctuations are small compared to the region that has been swept out
by the GRB and our finding that the circumburst environment can be
represented by a homogeneous medium is, on average, valid.  The
density fluctuations are between $\delta n / n \approx 0.3$ (for $n
\approx 0.1$ cm$^{-3}$) and $\approx 3$ (for $n \approx 10$
cm$^{-3}$).  Smaller regions will have larger density fluctuations
than larger regions will.  These fluctuations are similar to those
seen in the interstellar medium in our Galaxy \citep{DGR1989,FG2001}.


\section{Extinction in the Host\label{SECTION:extinct}}

     We can estimate the amount of extinction in the host galaxy along
the line of sight to \objectname{GRB~011211} by comparing the observed
spectral slope after correcting for Galactic reddening ($\beta = 0.61
\pm 0.15$) to the predicted intrinsic slope found in
\S~\ref{SECTION:medium} ($\beta_0 = 0.55 \pm 0.03$).  This difference
corresponds to $E_{B\!-\!V} = 0.02 \pm 0.05$ in the host galaxy, which
yields an extinction is $A_V = 0.06 \pm 0.15$ in the observer's frame.
Using the extinction law of \citet{CCM1989} we find $A_V \lesssim
0.03$ in the rest-frame of the burst.  This result depends somewhat on
the details of the extinction law used, but it suggests that there is
no significant extinction along the line of sight to
\objectname{GRB~011211} in its host galaxy.


\section{Conclusions\label{SECTION:conc}}
     
     We present early-time $V\!RI$ photometry and spectroscopy of the
OA of \objectname{GRB~011211} starting approximately 0.5 days after
the burst.  The spectrum contains several narrow metal lines that are
consistent with a redshift of $2.140 \pm 0.001$.  There is an
unidentified broad feature at $\approx 4600$ {\AA} that may be an
absorption system along the line of sight to the GRB.

     The OA is red with a spectral slope between 5505 {\AA} and 8060
{\AA} of $\beta = 0.61 \pm 0.15$ after correcting for Galactic
extinction, but ignoring extinction in the host.  This corresponds to
a color of $V\!-\!I = 0.70 \pm 0.13$.  The magnitude of the break in
the optical decay is consistent with the burst expanding into an
approximately homogeneous medium.  If we assume that GRBs have a the
standard energy suggested by \citet{FKS2001}, \citet{PKP2001}, and
\citet{PK2002} then the ambient medium near the burst has a particle
density of 0.1--10 cm$^{-3}$.  Comparing the observed color of the OA
with predictions of the standard beaming model suggests that the
rest-frame $V$-band extinction in the host galaxy is $A_V \lesssim
0.03$ mag.

     \objectname{GRB~011211} follows the same broad pattern
established by other long-duration GRBs.  The OA decays as a power law
with a slope of $\alpha = 0.83 \pm 0.04$ for the first $\approx 2$
days, and there is evidence for a break occurring between 1.52 and
2.72 days after the burst.  The slope of the light curve after the
break is $\alpha_2 \ge 1.4$.  We find evidence for variations of
$\approx 8$\% in the flux on time scales of 1--2 hours occurring
approximately half a day after the burst.  Interpreting this in the
framework of \citet{WL2000} suggests that there are density
fluctuations with scales of approximately 40--125 AU within $\approx
0.05$--0.20 pc of the GRB's progenitor.  The discovery of rapid
variations in the optical light from a GRB highlights the importance
of continuous, high-precision observations of GRBs at early times.


\acknowledgements

     We wish to thank the {\sl BeppoSAX\/} team, Scott Barthelmy, and
the GRB Coordinates Network (GCN) for rapidly providing precise GRB
positions to the astronomical community.  We also wish to thank Arne
Henden for providing precision photometry of stars in GRB fields.  The
authors would like to thank the anonymous referee for their helpful
comments on this paper.  STH and PMG acknowledge support from the NASA
LTSA grant NAG5--9364.  DB has been supported by NSF grand
AST--9979812.  This research has made use of the NASA/IPAC
Extragalactic Database (NED), which is operated by the Jet Propulsion
Laboratory, California Institute of Technology, under contract with
NASA\@.


  
\end{document}